\documentclass{PoS}
\usepackage[sort&compress]{natbib}
\bibpunct{[}{]}{,}{n}{}{}

\newcommand{\comment}[1]{}
\newcommand{\lr}[1]{ \left( #1 \right) }
\newcommand{\lrs}[1]{ \left[ #1 \right] }
\newcommand{\lrc}[1]{ \left\{ #1 \right\} }

\newcommand{\re}{ {\rm Re} \, }

\renewcommand{\det}[1]{ {\rm det} \left( #1 \right) }
\newcommand{\sign}{ {\rm sign} \,  }

\newcommand{\expa}[1]{ \exp{\left( #1 \right)} }

\title{Overlap Dirac operator with chiral chemical potential and Chiral Magnetic Effect on the lattice}

\ShortTitle{Dirac operator with chiral chemical potential and CME on the lattice}

\author{\speaker{P. V. Buividovich}\thanks{This work is dedicated to the memory of my Teacher, Prof.~Dr.~M.~I.~Polikarpov. This work was supported by the S.~Kowalewskaja award from the Alexander von Humboldt Foundation. I am grateful to Dr.~G.~Endrodi, Prof.~Dr.~A.~Sch\"{a}fer and to A.~Sadofyev for interesting discussions which motivated this work.}\\
        Institute for Theoretical Physics, Regensburg University, Universit\"{a}tsstrasse 31, D-93053 Regensburg, Germany\\
        E-mail: \email{pavel.buividovich@physik.uni-regensburg.de}}

\abstract{A self-consistent construction of the overlap lattice Dirac operator coupled to chiral chemical potential is proposed. With the help of the constructed operator we compute electric current induced by a constant magnetic field (Chiral Magnetic Effect). We find that the result disagrees with the one predicted by anomaly-based arguments and comment on the origin of this discrepancy. We demonstrate that a straightforward lattice calculation with a constant magnetic field and a uniform chiral chemical potential in fact corresponds to an infrared singularity in the dimensionally reduced polarization tensor and hence yields the result which is extremely sensitive to infrared regulators such as finite volume or finite temperature.}

\FullConference{31st International Symposium on Lattice Field Theory - LATTICE 2013\\
		July 29 - August 3, 2013\\
		Mainz, Germany}

\begin{document}
\sloppy

\section{Introduction}
\label{sec:intro}

 In these Proceedings we would like to check the validity of the well-known expression for the electric current generated due to the Chiral Magnetic Effect (CME) \cite{Kharzeev:08:2}
\begin{eqnarray}
\label{CMECurrentClassic}
 \vec{j} = \frac{\mu_5 \vec{B}}{2 \pi^2}
\end{eqnarray}
for lattice chiral fermions. The most obvious setup for lattice studies of the CME is the measurement of the electric current along constant magnetic field with quantized flux in a finite volume and in the presence of a constant chiral chemical potential \cite{Yamamoto:11:1}. While the way to introduce constant magnetic field on the lattice is well known, so far no systematic construction of the chiral lattice Dirac operator with chiral chemical potential was given in the literature. E.g. the results of \cite{Yamamoto:11:1} were obtained with the Wilson-Dirac fermions which are chirally symmetric only at small momenta. Since the CME current is saturated by fermionic states at high momenta close to the UV cutoff (see e.g. the derivation of \cite{Kharzeev:08:2} and also the discussion in Section \ref{sec:ir_sensitivity} below), it is reasonable to expect that the validity of the expression (\ref{CMECurrentClassic}) will crucially depend on whether the lattice Dirac operator has the (appropriately defined) chiral symmetry also at high momenta.

 In this work we construct the chiral lattice Dirac operator with finite chiral chemical potential $\mu_5$ basing on the overlap Dirac operator at finite chemical potential \cite{Neuberger:98:1, Bloch:06:1} and use it to calculate the CME current (\ref{CMECurrentClassic}) for free fermions in a constant magnetic field. It turns out that for symmetric lattices with equal spatial and temporal dimensions the result tends to the half of the current (\ref{CMECurrentClassic}) in the infinite volume limit. In order to explain the disagreement with the conventional expression (\ref{CMECurrentClassic}), we demonstrate that lattice measurements of the electric current (\ref{CMECurrentClassic}) in the presence of a homogeneous chiral chemical potential and a homogeneous external magnetic field in fact correspond to an infrared momentum-space singularity of the one-loop polarization tensor on the lowest Landau level. This singularity is regularized by infrared cutoffs such as finite volume and finite temperature. As a result, the relation (\ref{CMECurrentClassic}) is in general invalid on the lattice for constant $\vec{B}$ and $\mu_5$ and should only be realized if one considers the case of a spatially homogeneous chiral chemical potential which slowly varies in time.

\section{Overlap Dirac operator with chiral chemical potential}
\label{sec:overlap_construction}

 It is convenient to start the construction of the overlap Dirac operator with chiral chemical potential with the assumption that the chiral chemical potential $\mu_5$ can in general depend on spatial lattice coordinates. In any static background gauge field the derivative of the free energy of the fermion gas over $\mu_5\lr{x}$ should yield the static density of the axial charge $q_5\lr{x} = \bar{\psi}\lr{x} \gamma_0 \gamma_5 \psi\lr{x}$, which is, in turn, the imaginary part of the derivative of the free energy over the time-like component of the axial gauge field $A_{0}\lr{x}$. Taking into account that for the free fermion gas described by the Dirac operator $\mathcal{D}_{ov}$ the free energy is given by $\mathcal{F} = -T \log\mathcal{Z} = -T \log\det{ \mathcal{D}_{ov}}$, we get
\begin{eqnarray}
\label{ChemicalPotentialCouplingDef}
 -T \frac{\partial}{\partial \mu_5\lr{\vec{x}}} \log\mathcal{Z} = q_5\lr{\vec{x}}
 = -T \frac{\partial \, \log\det{ \mathcal{D}_{ov} }}{\partial \mu_5\lr{\vec{x}}}
 = i T \sum\limits_{\vec{x}, \tau} \frac{\partial \, \log\det{ \mathcal{D}_{ov}\lr{\mu_5} }}{\partial A_0\lr{x, \tau}} ,
\end{eqnarray}
where $T$ is the temperature and $\tau$ is the Euclidean time coordinate (we assume that the lattice spacing is equal to unity). Repeating this derivation for multiple derivatives over $\mu_5\lr{\vec{x}}$ at different spatial coordinates $\vec{x}$, we conclude that the chiral chemical potential should enter the overlap Dirac operator as the imaginary part of the time-like component $A_0\lr{x}$ of an axial gauge field $A_{\mu}\lr{x}$.

 We are thus led to a more general problem of constructing the overlap Dirac operator in the background of an axial gauge field. Continuum Dirac operator in the background of vector and axial gauge fields $V_{\mu}\lr{x}$ and $A_{\mu}\lr{x}$ reads
\begin{eqnarray}
\label{ContinuumDiracAxial}
 \mathcal{D}\lrs{V_{\mu}\lr{x}, A_{\mu}\lr{x}} = \gamma_{\mu}\lr{\frac{\partial}{\partial x^{\mu}} - i V_{\mu}\lr{x} - i \gamma_5 A_{\mu}\lr{x}} .
\end{eqnarray}
Under gauge transformations of both gauge fields $V_{\mu}\lr{x} \rightarrow V_{\mu}\lr{x} + \partial_{\mu} v\lr{x}$ and $A_{\mu}\lr{x} \rightarrow A_{\mu}\lr{x} + \partial_{\mu} a\lr{x}$, it transforms as follows:
\begin{eqnarray}
\label{ContinuumGaugeTransforms}
 \mathcal{D}\lrs{V_{\mu}\lr{x} + \partial_{\mu} v\lr{x}, A_{\mu}\lr{x} + \partial_{\mu} a\lr{x}}
 =
 e^{i v\lr{x} + i a\lr{x} \gamma_5} \,
 \mathcal{D}\lrs{V_{\mu}\lr{x}, A_{\mu}\lr{x}} \,
 e^{-i v\lr{x} + i a\lr{x} \gamma_5}  .
\end{eqnarray}
In this notation, we treat the local fields $v\lr{x}$ and $a\lr{x}$ as diagonal operators which act on some local field $\psi\lr{x}$ as $\lrs{v \psi}\lr{x} = v\lr{x} \psi\lr{x}$ and $\lrs{a \psi}\lr{x} = a\lr{x} \psi\lr{x}$. Note that while gauge transformations of the vector gauge field result in similarity transformations of the Dirac operator, this is not so for the gauge transformations of the axial gauge field. The fact that the Jacobian $\mathcal{J} \sim \det{e^{2 i a\lr{x} \gamma_5}}$ of the transformation (\ref{ContinuumGaugeTransforms}) is different from unity upon regularization is precisely the origin of the $U\lr{1}$ axial anomaly.

 The starting point of our construction of the chiral lattice Dirac operator with axial lattice gauge field $A_{\mu}\lr{x}$ is the lattice counterpart of the gauge transformations (\ref{ContinuumGaugeTransforms}). Following \cite{Kikukawa:98:1}, we replace the variation of the continuum Dirac operator (\ref{ContinuumDiracAxial}) $\delta_a \mathcal{D}\lrs{V_{\mu}\lr{x}, A_{\mu}\lr{x}}  = i \lrc{a\lr{x} \gamma_5, \mathcal{D}\lrs{V_{\mu}\lr{x}, A_{\mu}\lr{x}}}$ under infinitesimal gauge transformations $a\lr{x}$ of the axial gauge field with the local form of the L\"{u}scher transformations \cite{Luscher:98:1}:
\begin{eqnarray}
\label{LocalLuscherTransform}
 \delta_a \mathcal{D}_{ov} = \lr{1 - \mathcal{D}_{ov}/2} \gamma_5 a\lr{x} \mathcal{D}_{ov} + \mathcal{D}_{ov} a\lr{x} \gamma_5 \lr{1 - \mathcal{D}_{ov}/2} ,
\end{eqnarray}
where we have omitted the functional arguments of $\mathcal{D}_{ov}\lrs{V_{\mu}\lr{x}, A_{\mu}\lr{x}}$ for the sake of brevity. We have also assumed that both gauge fields $V_{\mu}\lr{x}$ and $A_{\mu}\lr{x}$ are associated with lattice links and are non-compact, so that $V_{-\mu}\lr{x} \equiv -V_{\mu}\lr{x - \hat{\mu}}$ and $A_{-\mu}\lr{x} \equiv -A_{\mu}\lr{x - \hat{\mu}}$. They are related to compact link variables via the standard exponentiation formula, e.g. $U_{\mu}\lr{x} = \expa{i V_{\mu}\lr{x}}$. We now require that the variation (\ref{LocalLuscherTransform}) is generated by an infinitesimal gauge transformation of the lattice axial gauge field $A_{\mu}\lr{x}$:
\begin{eqnarray}
\label{LatticeSmallAxialGaugeTransform}
  \delta_a \mathcal{D}_{ov}\lrs{V_{\mu}\lr{x}, A_{\mu}\lr{x}} = \mathcal{D}_{ov}\lrs{V_{\mu}\lr{x}, A_{\mu}\lr{x} + a\lr{x+\hat{\mu}} - a\lr{x}} - \mathcal{D}_{ov}\lrs{V_{\mu}\lr{x}, A_{\mu}\lr{x}} .
\end{eqnarray}
Expanding the right-hand side to the first order in $a\lr{x}$ and comparing the result with the infinitesimal variation due to a gauge transformation of the vector gauge field \cite{Kikukawa:98:1}, we obtain the following functional equation for $\mathcal{D}_{ov}\lrs{V_{\mu}\lr{x}, A_{\mu}\lr{x}}$:
\begin{eqnarray}
\label{LatticeDiracAxialDefiningEquation}
  \frac{\partial}{\partial A_{\mu}\lr{x}} \mathcal{D}_{ov}\lrs{V_{\mu}\lr{x}, A_{\mu}\lr{x}}
  =
  \frac{\partial}{\partial V_{\mu}\lr{x}} \mathcal{D}_{ov}\lrs{V_{\mu}\lr{x}, A_{\mu}\lr{x}}
  \gamma_5 \lr{1 - \mathcal{D}_{ov}\lrs{V_{\mu}\lr{x}, A_{\mu}\lr{x}}}
\end{eqnarray}
In order to solve this seemingly nonlinear equation, it is convenient to formulate it in terms of the projected overlap Dirac operator
\begin{eqnarray}
\label{ProjectedDiracDef}
 \tilde{\mathcal{D}}_{ov}\lrs{V_{\mu}\lr{x}, A_{\mu}\lr{x}} = \frac{2 \mathcal{D}_{ov}\lrs{V_{\mu}\lr{x}, A_{\mu}\lr{x}}}{2 - \mathcal{D}_{ov}\lrs{V_{\mu}\lr{x}, A_{\mu}\lr{x}}} .
\end{eqnarray}
This transformation projects the eigenvalues of $\mathcal{D}_{ov}$ from the Ginsparg-Wilson circle in the complex plane to the imaginary axis. Taking the derivative of (\ref{ProjectedDiracDef}) over the axial gauge field $A_{\mu}\lr{x}$, we get
\begin{eqnarray}
\label{ProjectedOperatorEquation}
  \frac{\partial}{\partial A_{\mu}\lr{x}} \, \tilde{\mathcal{D}}_{ov}
  =
  \frac{2}{2 - \mathcal{D}_{ov}} \,
  \frac{\partial}{\partial A_{\mu}\lr{x}} \, \mathcal{D}_{ov} \,
  \frac{2}{2 - \mathcal{D}_{ov}}
  = \nonumber \\ =
  \lr{1 + \tilde{\mathcal{D}}_{ov}/2} \frac{\partial}{\partial V_{\mu}\lr{x}} \, \mathcal{D}_{ov} \,
  \gamma_5 \lr{1 - \mathcal{D}_{ov}} \lr{1 + \tilde{\mathcal{D}}_{ov}/2}
  = \nonumber \\ =
  \lr{1 + \tilde{\mathcal{D}}_{ov}/2}
  \frac{\partial}{\partial V_{\mu}\lr{x}} \, \mathcal{D}_{ov} \,
  \lr{1 + \tilde{\mathcal{D}}_{ov}/2} \gamma_5
  =
  \frac{\partial}{\partial V_{\mu}\lr{x}} \, \tilde{\mathcal{D}}_{ov} \, \gamma_5  ,
\end{eqnarray}
where we have again omitted the arguments of $\mathcal{D}_{ov}$ and $\tilde{\mathcal{D}}_{ov}$. In the process of derivation we have used the Ginsparg-Wilson relations and the identity $\frac{2}{2 - \mathcal{D}_{ov}} = 1 + \frac{\tilde{\mathcal{D}}_{ov}}{2}$. We see that now the functional equation for $\tilde{\mathcal{D}}_{ov}$ is linear and is quite easy to solve explicitly. It is convenient to write the solution in terms of the chiral projectors $\mathcal{P}_{\pm} = \frac{1 \pm \gamma_5}{2}$:
\begin{eqnarray}
\label{ProjectedOperatorSolution}
 \tilde{\mathcal{D}}_{ov}\lrs{V_{\mu}\lr{x}, A_{\mu}\lr{x}} =
 \mathcal{P}_{-} \tilde{\mathcal{D}}_{ov}\lrs{V_{\mu}\lr{x} + A_{\mu}\lr{x}} \mathcal{P}_{+}
 +
 \mathcal{P}_{+} \tilde{\mathcal{D}}_{ov}\lrs{V_{\mu}\lr{x} - A_{\mu}\lr{x}} \mathcal{P}_{-}
\end{eqnarray}
where the projected overlap operators $\tilde{\mathcal{D}}_{ov}\lrs{V_{\mu} \pm A_{\mu}}$ are obtained from the overlap Dirac operators $\mathcal{D}_{ov}\lrs{V_{\mu} \pm A_{\mu}}$ in the background of the vector gauge fields $V_{\mu}\lr{x} \pm A_{\mu}\lr{x}$. The latter are coupled to fermions in a standard way by including the corresponding link factors into finite difference operators entering the Dirac-Wilson operator which is used to define the overlap Dirac operator \cite{Neuberger:98:1}.

 Returning now to the construction of the overlap Dirac operator at nonzero chiral chemical potential $\mu_5$, we conclude from equations (\ref{ProjectedOperatorSolution}) that the projected overlap operator in this case should take the following form:
\begin{eqnarray}
\label{ProjectedOperatorSolutionMu5}
 \tilde{\mathcal{D}}_{ov}\lr{\mu_5} =
 \mathcal{P}_{-} \tilde{\mathcal{D}}_{ov}\lr{\mu=+\mu_5} \mathcal{P}_{+}
 +
 \mathcal{P}_{+} \tilde{\mathcal{D}}_{ov}\lr{\mu=-\mu_5} \mathcal{P}_{-}
\end{eqnarray}
where $\tilde{\mathcal{D}}_{ov}\lr{\mu= \pm \mu_5}$ are the projected overlap operators at finite chemical potential which is equal to $+ \mu_5$ or $- \mu_5$. The corresponding overlap Dirac operators at finite chemical potential should satisfy the Ginsparg-Wilson relations and can be explicitly constructed following \cite{Bloch:06:1} as
\begin{eqnarray}
\label{ovdirac_finitemu}
 \mathcal{D}_{ov}\lr{\mu} = 1 + \gamma_5 U \sign{\re{\Lambda}} U^{-1} ,
\end{eqnarray}
where $\Lambda$ is the diagonal matrix of complex eigenvalues of the operator $\gamma_5 \mathcal{D}_w\lr{\mu}$, $\mathcal{D}_w\lr{\mu}$ is the Wilson-Dirac operator at finite chemical potential $\mu$ and $U$ is the similarity transformation which diagonalizes the operator $\gamma_5 \mathcal{D}_w\lr{\mu}$.

\section{Chiral Magnetic Effect with overlap Dirac operator}
\label{sec:CME}

 Using the overlap Dirac operator constructed in the previous Section, we now calculate the current (\ref{CMECurrentClassic}) for free fermions in the external magnetic field on a finite four-dimensional lattice. To this end we perform numerical diagonalization of the operator $\gamma_5 \mathcal{D}_w\lr{\mu}$ in the directions $x$ and $y$ perpendicular to the magnetic field $\vec{B} = B \vec{e}_z$ using \texttt{LAPACK}. In the directions parallel to the field the diagonalization is performed exactly in the plane wave basis $\psi\lr{\tau, z; k_{\tau}, k_z} = \sqrt{\frac{T}{L_z}} e^{i k_{\tau} \tau + i k_z z}$, where $k_{\tau} = 2 \pi T \lr{n + 1/2}$, $n \in \mathbb{Z}$ are the lattice Matsubara frequencies, $k_z = \frac{2 \pi m}{L_z}$, $m \in \mathbb{Z}$ is the $z$ component of the discrete lattice momenta and $L_z$ is the lattice size in the $z$ direction. The overlap Dirac operator at finite chemical potential $\mu = \pm \mu_5$ is computed using (\ref{ovdirac_finitemu}), and then the overlap Dirac operator at finite $\mu_5$ is computed from (\ref{ProjectedOperatorSolutionMu5}). In order to calculate the electric current $j_z$ in the direction of the magnetic field, we additionally introduce a constant vector gauge field $V_z$ in this direction. After the matrix of the operator $\mathcal{D}_{ov}\lr{\mu_5}$ is computed, we again perform its numerical diagonalization for the directions perpendicular to the magnetic field and calculate its eigenvalues $\lambda^{\lr{ov}}_i\lr{k_{\tau}, k_z}$. The current density is then calculated as the derivative of the free energy of the fermion gas over $V_z$:
\begin{eqnarray}
\label{CurrentDensityNumerical}
 j_z = \left. -\frac{T}{L_x L_y L_z} \frac{\partial}{\partial V_z} \log\det{\mathcal{D}_{ov}\lr{\mu_5}}\right|_{V_z = 0}
 = -\frac{T}{L_x L_y L_z} \sum\limits_{k_{\tau}, k_z}
 \sum\limits_i \frac{1}{\lambda^{\lr{ov}}_i\lr{k_{\tau}, k_z}} \frac{\partial \lambda^{\lr{ov}}_i\lr{k_{\tau}, k_z}}{\partial k_z} ,
\end{eqnarray}
where we have taken into account that the zero-frequency, zero-momentum component of $V_z$ simply amounts to the shift $k_z \rightarrow k_z - V_z$ in all the momentum sums.

\begin{figure*}
  \centering
  \includegraphics[width=5.2cm, angle=-90]{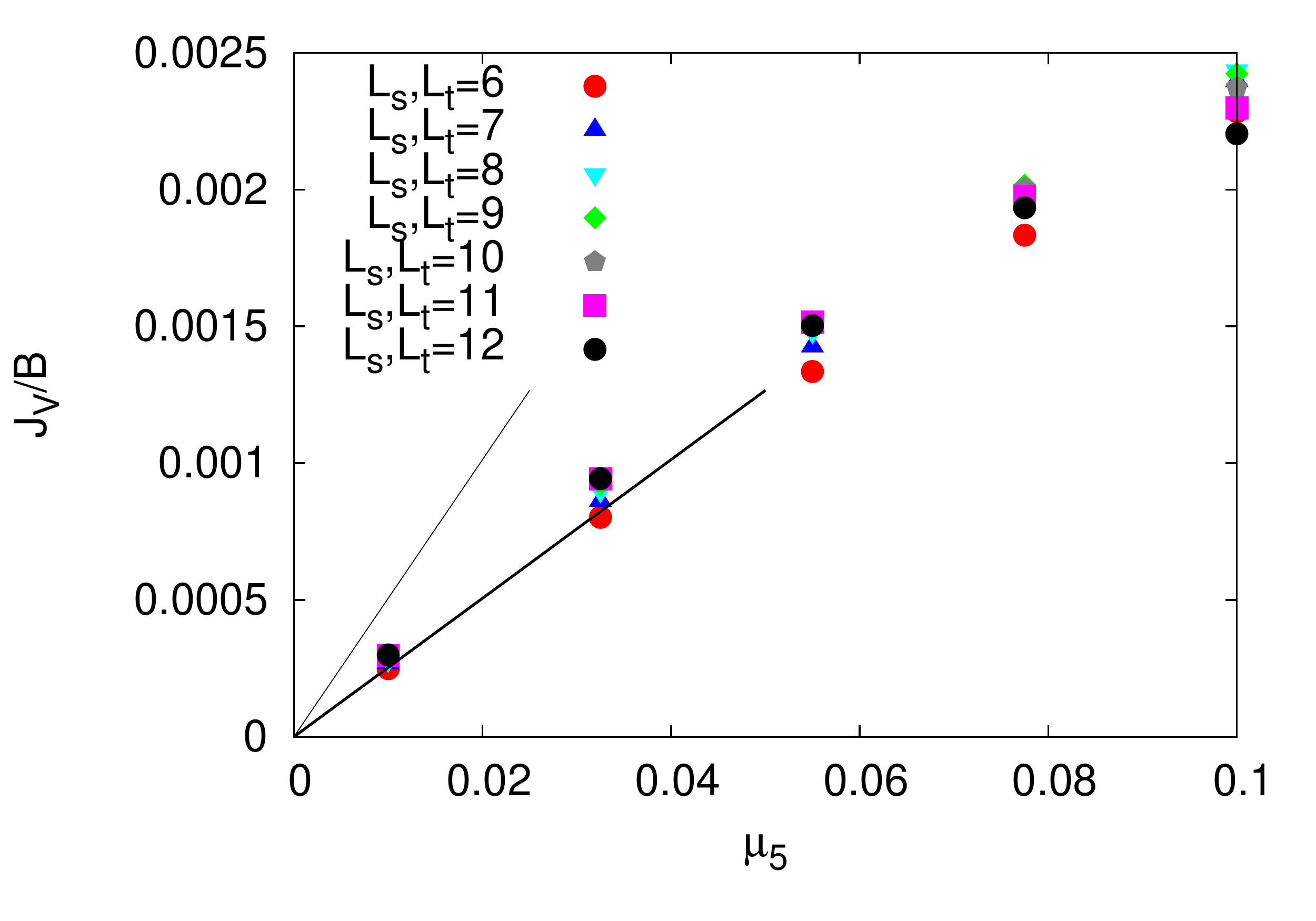}
  \includegraphics[width=5.2cm, angle=-90]{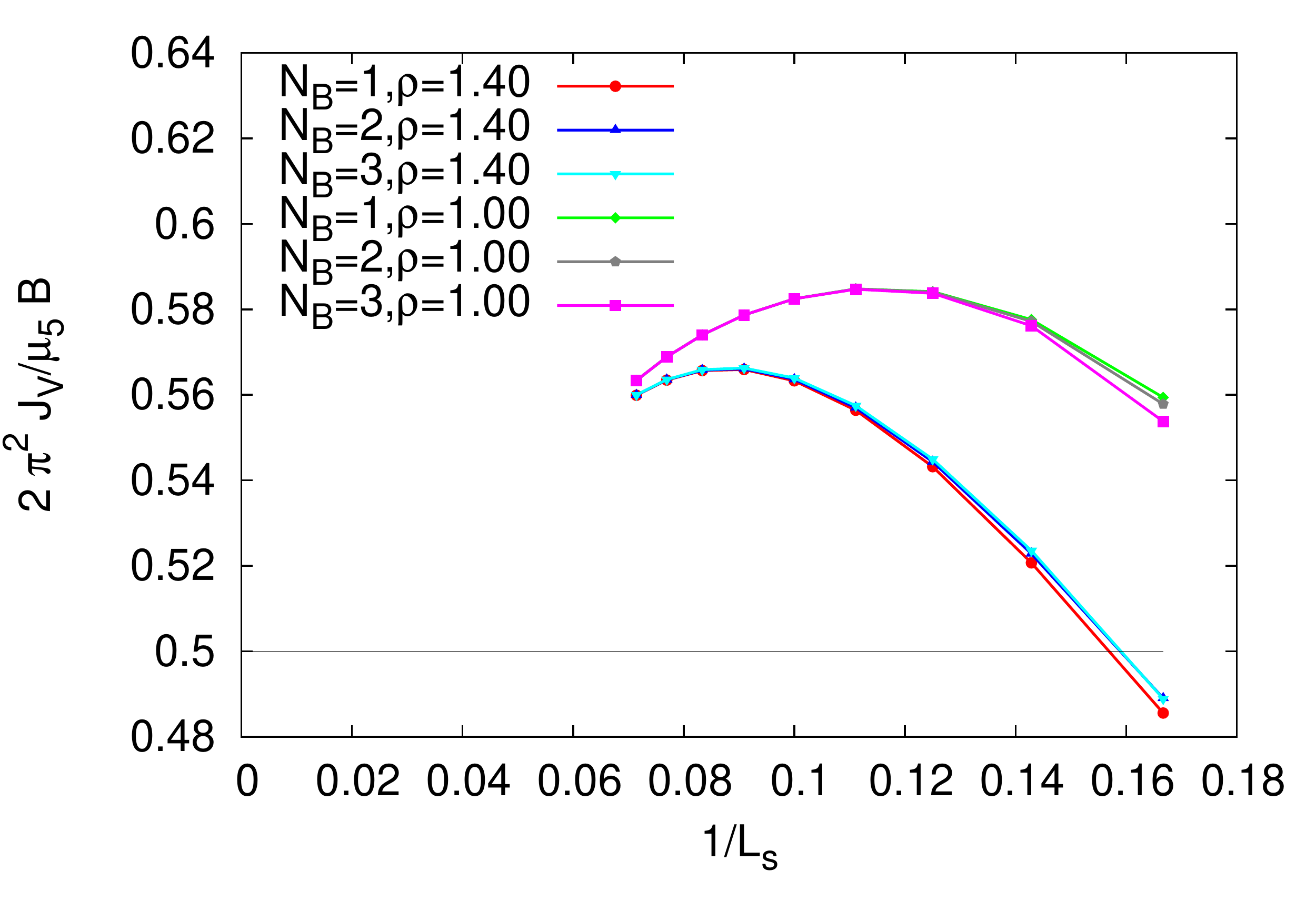}\\
  \caption{On the left: CME current (3.1) as a function of the chiral chemical potential $\mu_5$ for different lattice sizes and for one quantum of magnetic flux through the lattice section. Thin and thick solid lines correspond to the expression (1.1) and to the half of it, respectively. On the right: the coefficient $c$ in $j_z = \frac{c \mu_5 B}{2 \pi^2}$ as a function of inverse lattice size $L_s = L_t$ at different numbers $N_B$ of magnetic flux quanta and at different mass parameters $\rho$ in the overlap kernel $\gamma_5 \mathcal{D}_w$.}
  \label{fig:cme}
\end{figure*}

 The CME current calculated according to (\ref{CurrentDensityNumerical}) is illustrated on Fig. \ref{fig:cme}. On the left plot we show the ratio of the CME current to the magnetic field strength $B$ as a function of chiral chemical potential $\mu_5$ for different lattice sizes. The number of magnetic flux quanta $N_B = 2 \pi B/\lr{L_x \, L_y}$ through the lattice cross-section is fixed to be $N_B = 1$. Thin and thick straight solid black lines on the plot corresponds to the expression (\ref{CMECurrentClassic}) and to the half of it ($j_z = \frac{\mu_5 B}{4 \pi^2}$), respectively. One can see that the data is actually described by half of the conventional formula (\ref{CMECurrentClassic}) with a very good precision. In order to make this statement more precise, we perform the linear fits $j_z = \frac{c \, \mu_5 \, B}{2 \pi^2}$ of the dependence of $j_z/B$ on $\mu_5$. The coefficient $c$ is shown on the right plot on Fig. \ref{fig:cme} as a function of inverse lattice size $L_s$ (we assume $L_s = L_t$) for different values of $N_B$ and for different mass parameters $\rho$ in the overlap kernel $\gamma_5 \mathcal{D}_w$ \cite{Neuberger:98:1}. One can see that as $L_s$ tends to infinity, $c$ tends to $0.5$. Thus on the lattice we get only half of the CME current (\ref{CMECurrentClassic}), in disagreement with anomaly-based arguments \cite{Kharzeev:08:2}.

\section{Infrared sensitivity of the Chiral Magnetic Effect}
\label{sec:ir_sensitivity}

 In order to understand the origin of the discrepancy between the lattice results presented on Fig. \ref{fig:cme} and the conventional expression for the CME (\ref{CMECurrentClassic}), let us return for a while to the continuum theory. In a constant magnetic field the CME current is completely saturated by $N_B$-times degenerate states belonging to the lowest Landau level. These states are additionally labeled by the momenta $k_z$ in the direction of the magnetic field and the Matsubara frequencies $k_{\tau}$, so that the final expression for the CME current reads \cite{Kharzeev:08:2}:
\begin{eqnarray}
\label{CMECurrentDecomposed1}
 j_z = \frac{T B}{2 \pi L_z} \, \sum\limits_{k_{\tau}, k_z} \frac{2 \lr{k_z - \mu_5}}{k_{\tau}^2 + \lr{k_z - \mu_5}^2}  .
\end{eqnarray}
This expression is UV divergent and requires regularization (IR singularity of the propagator is removed since for anti-periodic boundary conditions $k_{\tau}$ is never zero). One possible way to obtain a finite result for the sum (\ref{CMECurrentDecomposed1}) is to sum over the Matsubara frequencies $k_{\tau}$ first (see e.g. \cite{Kharzeev:08:2}). In the limit of zero temperature and infinite $L_z$ we are then left with the following integral over $k_z$:
\begin{eqnarray}
\label{CMECurrentMomentumSpace}
 j_z=\frac{B}{\lr{2 \pi}^2} \, \int\limits_{-\infty}^{+\infty} dk_z \, \sign\lr{k_z - \mu_5} .
\end{eqnarray}
This is again a formally divergent integral, since the integrand does not tend to zero at $k_z \rightarrow \pm \infty$. Applying the most straightforward cutoff regularization, which amounts to replacing the infinite limits of integration with a finite interval $\lrs{-\Lambda_{UV}; \Lambda_{UV}}$ with $\Lambda_{UV} \gg \mu_5, \, \sqrt{B}$, we obtain exactly the expression (\ref{CMECurrentClassic}) \cite{Kharzeev:08:2}. However, a closer inspection reveals that such a cutoff regularization in fact violates the conservation of vector current and is thus not physically consistent.

 To regularize the expression (\ref{CMECurrentDecomposed1}) in a consistent way, we first note it is proportional to the expectation value of the electric current $j_z$ for a two-dimensional fermions restricted to the $z-t$ plane in the four-dimensional space in the presence of a static space-independent vector gauge field $V_z \equiv \mu_5$. Assuming that $V_z \equiv \mu_5$ is small, we can expand (\ref{CMECurrentDecomposed1}) to the linear order in $\mu_5$. We then conclude that the CME current is proportional to the $zz$ component of the one-loop polarization tensor $\Pi_{ij}\lr{q_{\tau}, q_z}$ at zero photon frequency $q_{\tau} = 0$ and zero photon momentum $q_z = 0$ in 2D QED:
\begin{eqnarray}
\label{CMECurrentPolarizationTensor}
 j_z = \frac{B}{2 \pi} \Pi_{zz}\lr{q_{\tau}=0, q_z=0} \mu_5 .
\end{eqnarray}
In a regularization consistent with vector current conservation, $\Pi_{ij}\lr{q}$ should have the form (see e.g. \cite{Chen:99:1}) $\Pi_{ij}\lr{q} = \frac{1}{\pi} \frac{\delta_{ij} q^2 - q_i q_j}{q^2}$, thus
\begin{eqnarray}
\label{PolarizationTensorZZ}
 \Pi_{zz}\lr{q} = \frac{1}{\pi} \frac{q_{\tau}^2}{q_z^2 + q_{\tau}^2} .
\end{eqnarray}
We see now that the point $q_{\tau}=0$, $q_z=0$ is a singular point of the function (\ref{PolarizationTensorZZ}), so that the value of the function crucially depends on the limit in which this singular point is approached. If we first take the limit of zero frequency $q_{\tau}$ and then send $q_z$ to zero, we get identically zero. This corresponds to the situation of time-independent $\mu_5$ which very slowly varies in space. On the other hand, if we first set $q_z = 0$ and only then set $q_{\tau} \rightarrow 0$, we recover the conventional answer (\ref{CMECurrentClassic}). This limit corresponds to spatially constant $\mu_5$ which slowly changes with time.

 On the lattice both $q_{\tau}$ and $q_z$ become discrete variables, so it does not make sense to distinguish the order of limits $q_{\tau} \rightarrow 0$ and $q_z \rightarrow 0$. From the lattice results presented above we see that instead the finite lattice size acts as an infrared cutoff which regularizes the IR singularity in (\ref{PolarizationTensorZZ}) and replaces the indefinite continuum result (\ref{PolarizationTensorZZ}) with the mean of the two limits discussed above. This is then exactly half of the expression (\ref{CMECurrentClassic}).

 Another, more technical argument which demonstrates IR sensitivity of the CME current (\ref{CMECurrentClassic}) can be given based on the expression (\ref{CurrentDensityNumerical}). It can be represented as a sum over $k_{\tau}$ and $k_z$ of the derivative of the function $\mathcal{F}\lr{k_{\tau}, k_z} = \prod\limits_i \lambda^{\lr{ov}}_i\lr{k_{\tau}, k_z}$ over $k_z$. Now if we fix the temperature $T$ in lattice units and tend $L_z$ to infinity, the sum over $k_z$ turns into an integral of the form $\int dk_z \, \partial \mathcal{F}\lr{k_{\tau}, k_z}/\partial k_z$, which is clearly zero on the lattice due to compactness and periodicity of the Brillouin zones. We again see that the final result for the CME current strongly depends on the IR regularization - in this case, on the order in which the limits of the infinite spatial volume and zero temperature are taken.

\section{Conclusions}
\label{sec:conclusions}

 We conclude that the most straightforward way to measure CME on the lattice by measuring the electric current in the limit of static and spatially homogeneous magnetic field and chiral chemical potential corresponds to an infrared singularity of the one-loop polarization tensor in 2D QED. For free fermions on a finite lattice this singularity is resolved in such a way that exactly half of the conventional CME current (\ref{CMECurrentClassic}) is observed. These observations imply, in particular, that the CME current (\ref{CMECurrentClassic}) is not in general robust against IR effects and one can expect some corrections due to interactions between fermions.


\end{document}